\documentclass[aps,pre,showpacs,twocolumn,superscriptaddress,floatfix]{revtex4}
\usepackage{graphicx}
\usepackage{epsfig} 
\usepackage{epstopdf}
\usepackage{color}
\usepackage{amsfonts}
\usepackage{amssymb,amsmath}
\usepackage{bbold}

\begin{document}
 
\title{Importance sampling with imperfect cloning for the computation of 
			generalized Lyapunov exponents}

\author{Celia Anteneodo}
\email{celia.fis@puc-rio.br}
\author{Sabrina Camargo}
\email{sabrina.camargo@ufabc.edu.br}
\affiliation{Department of Physics, PUC-Rio, Caixa Postal 38071,
             22452-970 Rio de Janeiro, Brazil} 
\author{Ra\'ul O. Vallejos}
\email{vallejos@cbpf.br}
\affiliation{Centro Brasileiro de Pesquisas F\'{\i}sicas (CBPF), Rua Dr.~Xavier Sigaud 150, 
             22290-180 Rio de Janeiro, Brazil}

\date{\today}

\begin{abstract}
We revisit the numerical calculation of generalized Lyapunov exponents, $L(q)$, 
in deterministic dynamical systems.
The standard method consists of adding noise to the dynamics in order to use
importance sampling algorithms.  
Then $L(q)$ is obtained by taking the limit noise-amplitude $\to$ 0 after the calculation.
We focus in a particular method that involves periodic cloning and pruning of a set of
trajectories. 
However, instead of considering a noisy dynamics, we implement an imperfect (noisy) cloning.
This alternative method is compared with the standard one and, when possible, with analytical 
results.  
We use as workbench the asymmetric tent map, the standard map, and a system of coupled symplectic maps.
The general conclusion of this study is that the imperfect-cloning method performs as well 
as the standard one, with the advantage of preserving the deterministic dynamics.
\end{abstract}


\maketitle

\section{Introduction}
\label{sec:intro}

The (maximal) Lyapunov exponent $\lambda$ measures the sensitivity to infinitesimal perturbations
in dynamical systems. 
Its definition involves the infinite-time limit \cite{pikovsky16,ott93}
\begin{equation}
\lambda=\lim_{t\to\infty} \frac{1}{t} \ln  \frac{\left|\delta {\bf x}(t) \right|}{\left|\delta {\bf x}_0 \right|}   \,.
\label{eq:lambda}
\end{equation}
Here $\delta {\bf x}$ represents the distance vector between two infinitesimally close orbits (tangent vector).
If $\lambda>0$ the perturbation grows exponentially fast and the system is chaotic, unpredictable.
The Lyapunov exponent does not depend on the initial condition ${\bf x}_0$, provided ${\bf x}_0$
lies within a connected region of phase space.

However, if propagation over finite times is considered, e.g., for the sake of numerical calculations, then the
finite-time Lyapunov exponent $\lambda(t)$ fluctuates with initial conditions. 
In this case a thorough assessment of predictability requires the consideration of the full distribution of 
$\lambda(t)$, i.e., $P(\lambda(t))$. 
In order to characterize finite-time fluctuactions, one can compute different moments $\left|\delta {\bf x}(t) \right|^q$
and introduce the so-called (maximal) generalized Lyapunov exponents of order $q$:
\cite{fujisaka83,benzi85,crisanti93,castiglione08}
\begin{equation}
L(q) = \lim_{t\to\infty} \frac{1}{q t} 
        \ln \left\langle   \frac{\left|\delta {\bf x}(t) \right|^q}{\left|\delta {\bf x}_0 \right|^q} \right\rangle \, ,
\label{eq:Lq}
\end{equation}
where brackets indicate average over initial conditions (according to the invariant measure).
The usual Lyapunov exponent can be obtained as \cite{zillmer03,pikovsky16}
\begin{equation}
\lambda = \lim_{q \to 0} L(q) \,.
\end{equation}
The connection between $L(q)$ and $P(\lambda(t))$ can be easily seen.
In fact, we have
\begin{equation}
\left\langle  \frac{\left|\delta {\bf x}(t) \right|^q}{\left|\delta {\bf x}_0 \right|^q} \right\rangle =
\int d\lambda(t) \, P(\lambda(t)) \, e^{q t \lambda(t)} 
\end{equation}
[see Eq.~(\ref{eq:Lq})].

Generalized Lyapunov exponents appear in many problems, e.g., 
characterization of intermittency 
\cite{fujisaka83,benzi85,behnia12}, 
Anderson localization 
\cite{paladin87,kuzovkov02,gurevich11,zilly12}, 
transport, mixing and reaction of constituents in complex fluid flows 
\cite{vanneste10,haynes14,kahlen17}, etc.
We can think of intermittency as a non-uniform distribution in time of ``chaotic
behavior" \cite{benzi85}, which can be described by $P(\lambda(t))$, or, 
equivalently, by $L(q)$.
The transfer matrix method establishes a link between temporal intermittency and the
properties of spatial decay of the wave function in one
dimensional disordered systems (Anderson localization).
The family of localization lengths $\xi_q$, introduced in \cite{paladin87}, 
and related to the decay of correlation functions of different order,
correspond to generalized Lyapunov exponents in the transfer matrix language. 
Concerning complex fluids, the generalized Lyapunov exponents associated
with the stretching have been found to control the decay rate
of purely advected passive scalars \cite{tsang05,haynes05}.

When $q$ is large enough, the average in (\ref{eq:Lq}) is dominated by rare
events, i.e., trajectories having finite-time Lyapunov exponents
far away from the average value.
So, in general, standard sampling methods produce wrong results.
These numerical difficulties involved in the calculation of $L(q)$ are well known 
\cite{tailleur07,castiglione08,anteneodo10,vanneste10,pikovsky16}. 
The way out is using Monte Carlo importance sampling methods \cite{liu01}.

Recently Vanneste proposed and tested one such method for random maps.
This is a Monte Carlo algorithm involving periodic cloning/pruning
steps that select those trajectories which most contribute to $L(q)$ \cite{vanneste10}. 
This algorithm is a variant of that developed by Tailleur and Kurchan
for selecting trajectories with unusual $\lambda$ \cite{tailleur07} 
(see also \cite{tailleur07b,laffargue13}). 
These algorithms can be traced back to the go-with-the-winners methods discussed
by Grassberger \cite{grassberger02}.

In the case of weak intermittency and/or small $q$,
one can use the expansion in the cumulants of 
the distribution of $\lambda(t)$: \cite{castiglione08,anteneodo10}
\begin{equation}
L(q) \sim \sum_{n\ge 1} \frac{ (qt)^{n-1} }{n!} \kappa_n(t)  \,,
\label{eq:Lqcum}
\end{equation}
where $\kappa_n$ are the $n$th-order cumulants of $P(\lambda(t))$.
In numerical calculations one usually considers the first few cumulants.
Both propagation time and number of trajectories must be large
enough for the required cumulants set to a well defined value \cite{anteneodo10}. 

The importance sampling methods mentioned above either consider a noisy system
\cite{vanneste10} or add noise to an otherwise deterministic dynamics \cite{tailleur07}.
In the present paper we focus on the numerical calculation of $L(q)$ for
deterministic dynamical systems. 
However, instead of considering a noisy dynamics, we implement an imperfect (noisy) cloning
\cite{tailleur07b,pikovsky16}.
This alternative method is compared with the standard one and, when possible, with analytical 
results.  
We use as workbench 
the asymmetric tent map \cite{castiglione08}, 
the standard map \cite{ott93}, 
and a system of coupled symplectic maps \cite{crisanti88,falcioni91,kuptsov11,manchein12}.
The general conclusion of this study is that the imperfect-cloning method performs as well 
as the standard one, 
with the advantage of preserving the dynamics.

The rest of the paper is organized as follows.
In Sec.~\ref{sec:num} we describe the numerical methods. 
Then we proceed with the applications to our three model systems
(Secs.~\ref{sec:tentmap} to \ref{sec:coupledmaps}).
Finally, in Sec.~\ref{sec:conclusions} we discuss our results.

%%%%%%%%%%%%%%%%%%%%%%%%%%%%%%%%%%%%%%%%%%%
\section{Numerical methods} 
\label{sec:num}

The Benettin method for calculating the usual Lyapunov exponent 
$\lambda$ [Eq.~(\ref{eq:lambda})]
relies on the propagation of $K$ pairs of trajectories, 
and approximating tangent vectors by finite distance vectors between trajectories. 
Let us call the total propagation time $N$ (we are dealing with maps, then time is discrete).
As these vectors must remain small for the linear approximation to be valid, 
they are periodically renormalized \cite{benettin80}.
Thus, after a time $N$, one has $K$ finite-time Lyapunov exponents, 
i.e., a distribution $P(\lambda(N))$. 
By averaging over this distribution, one obtains the estimate for $\lambda$.

As suggested by Eq.~(\ref{eq:Lq}), Bennetin's method can also be used for computing
$\left|\delta {\bf x}(N) \right|^q$, and, after averaging, etc., one would have an estimate
for $L(q)$. 
However, this simple averaging can lead to wrong results, especially
in case of large $q$ and/or strong intermittency. 
This method will be referred to as {\em brute-force} Monte Carlo sampling \cite{vanneste10}
and its result as $L_{BF}(q)$.

One possible way of improving brute-force sampling is to use the cumulant expansion (\ref{eq:Lqcum}) 
\cite{anteneodo10,vallejos12}.
For instance, truncation at second order (Gaussian approximation) gives:
\begin{equation}
L_G(q) \equiv  \kappa_1 + \frac{qN}{2} \kappa_2 \, ,
\label{eq:Lgauss}
\end{equation}
where $\kappa_1 \equiv \lambda$ and $\kappa_2$ are, respectively 
the average and the variance of $P(\lambda(N))$ \cite{zillmer03,xie09,kuptsov11,pikovsky16}.
In chaotic systems, the calculation of the second cumulant $\kappa_2$ offers no problem: 
it stabilizes relatively fast at a definite value (plus small fluctuations). 
On the contrary, higher cumulants, e.g., $\kappa_3$ and $\kappa_4$, 
being very sensitive to the tails of $P(\lambda(N))$, are much trickier \cite{anteneodo10,vallejos12}.
However, if phase space is mixed, even the calculation of the variance $\kappa_2$ may be problematic;
see Secs.~\ref{sec:stdmap} and \ref{sec:coupledmaps}.

%%%%%%%%%%%%%%%%%%%%%%%%%%%%%%%%%
\subsection{Importance sampling} 
\label{sec:ip}

We begin by describing
succinctly Vanneste's importance sampling algorithm 
to calculate $L(q)$ \cite{vanneste10},
focusing only on those aspects that are relevant for the present paper. 

The algorithm starts using the Bennetin method, i.e., we launch $K$ pairs of close trajectories,
initial conditions chosen at random, and distances fixed to a common value, i.e., 
$|\delta {\bf x}_k(0)|=d_0 \ll 1$, for $1 \le k \le K$.
We let the trajectories evolve according to the map dynamics,
and follow in time the distances $\delta {\bf x}_k(n)$ up to a given time $\Delta_{res}$.
Assume that each pair of trajectories is labeled by their distance $\delta {\bf x}_k$.
Now resample according to:
\begin{equation}
\delta {\bf x}_k = \delta {\bf x}_{J} \, ,
\label{eq:res}
\end{equation}
where $J$ is a random variable taking values in $\{1, \ldots ,K \}$ with probability
\begin{equation}
P(J=j) = \frac{\alpha_j}{\beta} \, .
\label{eq:prob}
\end{equation}
Here we have defined 
\begin{equation}
\alpha_k =  \left|\delta {\bf x}_k \right|^q \, ,
\label{eq:alpha}
\end{equation}
and
\begin{equation}
\beta =  \sum_{k=1}^K \left|\delta {\bf x}_k \right|^q \, .
\label{eq:beta}
\end{equation}

It is possible to use other resampling schemes \cite{grassberger02,tailleur07,tailleur07b,pikovsky16}.
Whether these alternatives are helpful will depend on the problem at hand \cite{grassberger02}.
Thus, in addition to Vanneste's scheme, 
we chose to try also the cloning/pruning strategy used by Tailleur in \cite{tailleur07b}.
According to this strategy at each resampling step, 
each $\delta {\bf x}_k$ is replaced by $\tau$ clones, 
where $\tau$ is a random integer defined by 
\begin{equation}
\tau =  \left\lfloor  K \frac{\alpha_j}{\beta} + \epsilon \right\rfloor  \, ,
\end{equation}
$\epsilon$ being a random number uniformly distributed in $[0,1]$.
If $\tau=0$, then the pair of trajectories characterized by $\delta {\bf x}_k$ is killed. 
If $\tau>1$, then $\tau-1$ clones are created.
After this replication phase the number of trajectories may have changed. 
Let's call the difference $\Delta K$. 
If $\Delta K > 0$ or $\Delta K < 0$, then $\Delta K$ trajectories are respectively killed or cloned randomly. 
Thus, we keep the number of trajectories fixed ($=K$) \cite{tailleur07b}.

So, pairs of trajectories are cloned or pruned according with 
the schemes described above.
After this resampling step, distance-vectors are normalized to 
equal moduli, and evolution resumes. 
In a noisy dynamics, the clones do spread,
and after each time interval $\Delta_{res}$ a new resampling is made. 
The algorithm continues alternating between free propagation  
and resampling until time $N$.
Each resampling step produces a sum $\beta$ (\ref{eq:beta}).
Finally, the generalized exponent $L(q)$ is calculated from all the $\beta$s: \cite{vanneste10}
\begin{equation}
L(q) = \frac{1}{q M} \log \frac{1}{K^M} \beta_1 \beta_2 \ldots \beta_M \, ,
\end{equation}
where $M$ is the number of resampling steps.

When the dynamics is deterministic, the natural trick is to add some noise to
the equations of motion (say, of amplitude $\eta$), use the algorithm described above
to calculate $L(q,\eta)$, and then make $\eta \to 0$ \cite{tailleur07}.
There is, however, the simpler alternative of preserving the determinism of the dynamics 
but introducing noise immediately after the resampling step. 
The result of this process is that clones are no more identical.
We will call this modification of the algorithm {\em imperfect cloning}.
We implement imperfect cloning by just adding some noise 
to all the trajectories, i.e.,  
\begin{equation}
x_k = x_k + \xi_k  \, ,
\end{equation}
where $\xi_k$ are independent random variables uniformly distributed in $[-\eta,\eta]$
(we are considering a one-dimensional map; for other maps, see below).
Both trajectories in a Benettin pair $\{x_k,x_k+ \delta x_k\}$ suffer the same noise,
so, the distance vectors $\delta x_k$ are not affected by the imperfect cloning.

In the forthcoming sections we test the method in several model systems, 
trying to determine the best ranges for the parameters: 
$K$ (number of samples), 
$N$ (propagation time), 
$\eta$ (noise amplitude, either dynamical or for imperfect cloning),
$\Delta_{res}$ (period of resampling). 

A few general criteria for choosing parameter values can be given a priori.
The renormalization period for the Benettin method \cite{benettin80}, $\Delta_{ren}$, 
should be of the order of the Lyapunov time $\equiv 1/\lambda$.
The resampling time, $\Delta_{res}$, must be large enough in order to allow spreading
of clones.
Vanneste has argued that the condition for the validity of 
the importance sampling method is $K \gg N$ \cite{vanneste10}.

For the tent map and the standard map we chose $q$ large enough ($q=8$), 
so that importance sampling would be essential to obtain correct results.
Brute force sampling and the Gaussian approximation are bound to fail
in this case. So, $q=8$ constitutes a very stringent test for our method.

%%%%%%%%%%%%%%%%%%%%%%%%%%%%%%%%%%%%%%%%%%%
\section{Tent map} 
\label{sec:tentmap}

The asymmetric tent map is defined by:
 \begin{equation}
   x_{n+1}   = \left\{ \begin{array}{cl}
           \dfrac{x_n }{a} & \mbox{for $0 \le x \le a$} \,, \vspace{0.5pc} \\
        \dfrac{1-x_n}{1-a} & \mbox{for $a  <  x \le 1$} \,.
				               \end{array} 
							\right.                         
   \label{eq:tent}
 \end{equation}
The asymmetry parameter will be set to $a=0.3$. 
The simplicity of this map permits the analytical calculation of the 
generalized Lyapunov exponents: \cite{castiglione08}
\begin{equation}
L(q)= \log [a^{1-q} + (1-a)^{1-q}]/q \,.
\label{eq:tent-teo}
\end{equation}
The Gaussian approximation is just the linear expansion of $L(q)$ about $q=0$.
\begin{figure}[b]
\centering
\includegraphics[scale=0.95]{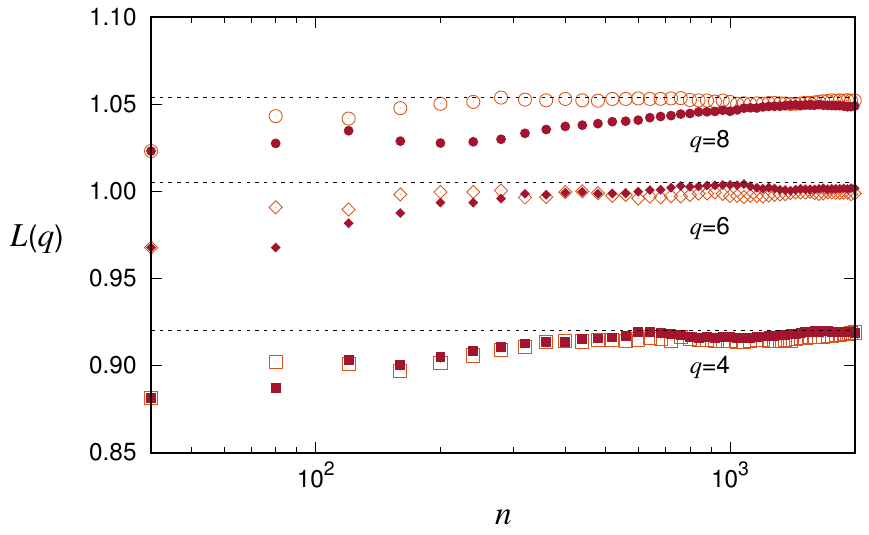} \\ 
\caption{
Tent map: generalized exponents $L(q)$ ($q=4,6,8$) versus propagation time $n$.
Dashed lines correspond to theoretical values.
We show the results of both importance sampling algorithms described 
in Sec.~\ref{sec:ip}: Tailleur's (hollow symbols) and Vanneste's (full, smaller, symbols). 
Cloning noise was set to $\eta=10^{-5}$. The number of samples is $K=1000$.
}
\label{fig:comp}
\end{figure}

In Fig.~\ref{fig:comp} we show a comparison of both resampling schemes described
in Sec.~\ref{sec:ip} combined with imperfect cloning.
The first observation is that both numerical methods coincide with the theoretical 
predictions for large times, i.e., both methods are equally accurate. 
However, Tailleur's resampling is faster, at least for $q=6,8$. 
We checked that this behavior persists for all the systems we tested, i.e., both
methods are equally accurate, but Tailleur's is equally fast or faster. 
So, we decided to use Tailleur's cloning/pruning scheme in the rest of the paper.
We remark: it is a question of speed, not accuracy.

\begin{figure}[ht]
\centering
\includegraphics[scale=0.95]{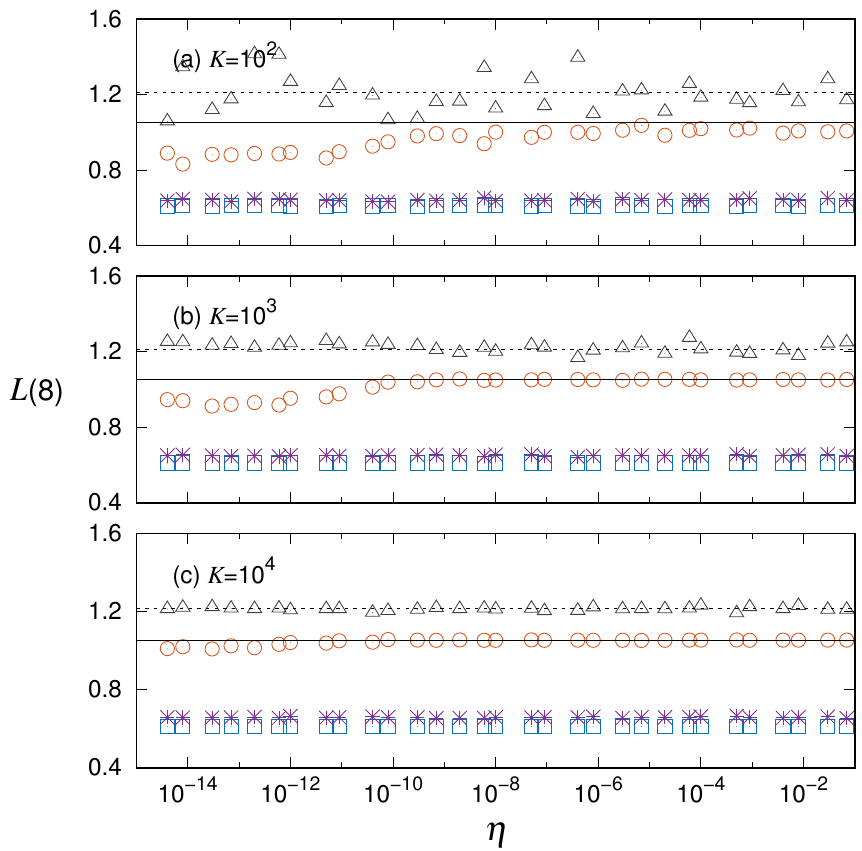} \\ 
\caption{Tent map: generalized exponent $L(8)$ 
versus noise amplitude $\eta$ of the imperfect cloning process (circles).
Shown are also the Gaussian approximation $L_G(8)$ (triangles), 
the results of the brute-force method $L_{BF}$ (stars) and 
the maximal Lyapunov exponent $\lambda$ (squares).
Lines correspond to theoretical values.
Three different sample-sizes were used:  $K=10^2$ (a), $10^3$ (b) and $10^4$ (c).
The propagation time for all the calculations is $N=10^3$.
 }
\label{fig:lyap_tent_q8_clon}
\end{figure}

In Fig.~\ref{fig:lyap_tent_q8_clon} we show the results for $L(q=8)$
using our algorithm of imperfect cloning.
This is a case of large intermittency, suitable to test our method.
The method performs very well, except for smallest sample size, i.e., $K=100$.
This is consistent with the validity criterion $K \gg N$.  
The calculation is insensitive to the noise amplitude provided it is not too
small, i.e., $\eta \gtrsim 10^{-10}$.
It is clear that both the Gaussian approximation and brute-force sampling give wrong results.

The values of $L(q)$ obtained through the Gaussian approximation and 
brute-force sampling do not depend on the noise amplitude $\eta$,
given that the noise only acts on the cloning procedure (the same is true for $\lambda$). 
The fluctuations observed in $L_G$, $L_{BF}$, and $\lambda$  
are due to the use of a different set of initial conditions for each value of $\eta$
(Fig.~\ref{fig:lyap_tent_q8_clon}).

\begin{figure}[ht]
\centering
\includegraphics[scale=0.95]{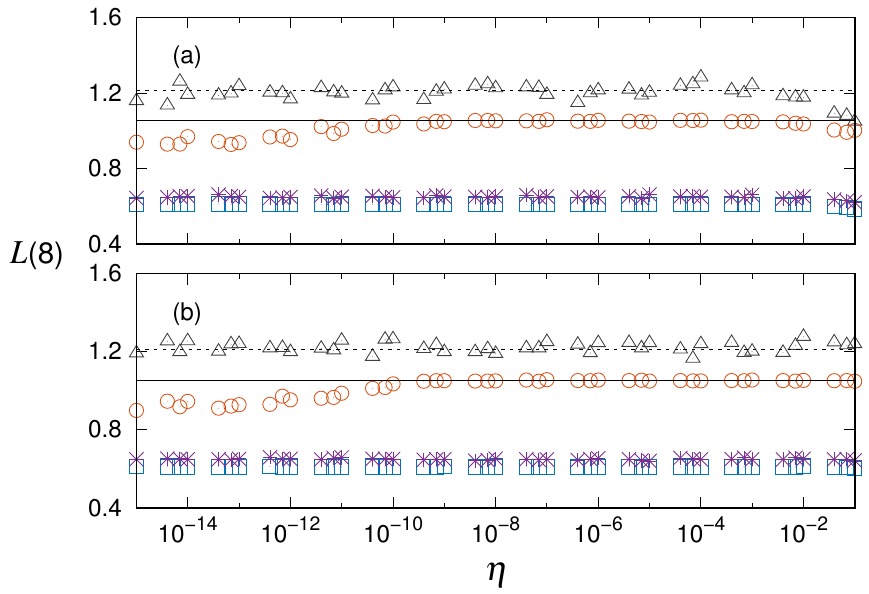}\\ 
\caption{ Tent map: generalized exponent $L(8)$  
versus the amplitude $\eta$ of the dynamical noise (circles). 
We also show the Gaussian approximation (triangles), 
brute-force sampling $L_{BF}$ (stars) and 
the maximal Lyapunov exponent (squares). 
In all cases $K=10^3$ and $N=10^3$. 
The dynamical noise is 
(a) additive, that is, added to the variable of state $x$, or 
(b) multiplicative, that is, to the map control parameter $a$. 
Lines correspond to theoretical values. 
}
\label{fig:lyap_tent_perturbed}
\end{figure}

Figure~\ref{fig:lyap_tent_perturbed} exhibits also $L(q=8)$ but calculated according 
to the standard method, i.e., by adding noise to the mapping and using perfect cloning.
Two cases were considered: 
(i)  noise is added to the state variable $x$, and 
(ii) noise is added to the map parameter $a$ (which amounts to multiplicative noise in $x$).
This figure presents an analogous behavior to Fig.~\ref{fig:lyap_tent_q8_clon}.
The noisy-dynamics calculation reproduces the analytical value of $L(8)$
and is also independent of the noise amplitude (within certain bounds).

We used the following parameter values: $d_0=10^{-12}$, $\Delta_{ren}=4$,
and $\Delta_{res}=40$.

%%%%%%%%%%%%%%%%%%%%%%%%%%%%%%%%%%%%%%%%%%%		
\section{Chirikov standard map} 
\label{sec:stdmap}		
		
The standard map is a two dimensional symplectic system defined by the equations
\begin{eqnarray}
   p_{n+1}    &=&  p_n - \mathsf{K} \sin q_n \,, \\
	 q_{n+1}    &=&  q_n + p_{n+1}  \,,
 \end{eqnarray}	 
where both variables, $q$ and $p$, are taken modulo $2\pi$.
The parameter $\mathsf{K}$ controls the map's chaoticity. 
For $\mathsf{K} \gtrsim 7$ the phase space appears to be covered by a single
chaotic sea, however islets of regularity do exist for arbitrarily large values 
of $\mathsf{K}$ \cite{chirikov79,lichtenberg92}. 
As $\mathsf{K}$ decreases the area filled with islands increases. 
For small $\mathsf{K}$, e.g., $\mathsf{K} \approx 2$, the phase portrait has a very rich structure \cite{lichtenberg92,ott93}.  

For the standard map there are neither analytical nor numerical results for $L(q)$ (to the best of our knowledge),
except an approximate formula for $\lambda$, valid for $\mathsf{K} \gtrsim 6$ \cite{chirikov79}:
\begin{equation}
\lambda \approx \log \frac{\mathsf{K}}{2} \,.
\label{eq:lambda_app}
\end{equation}
Tomsovic and Lakshmirayan have improved the formula above and provided approximate 
expressions for higher cumulants of $P(\lambda)$ \cite{tomsovic07}.

\begin{figure}[ht]
\centering
\includegraphics[scale=0.95]{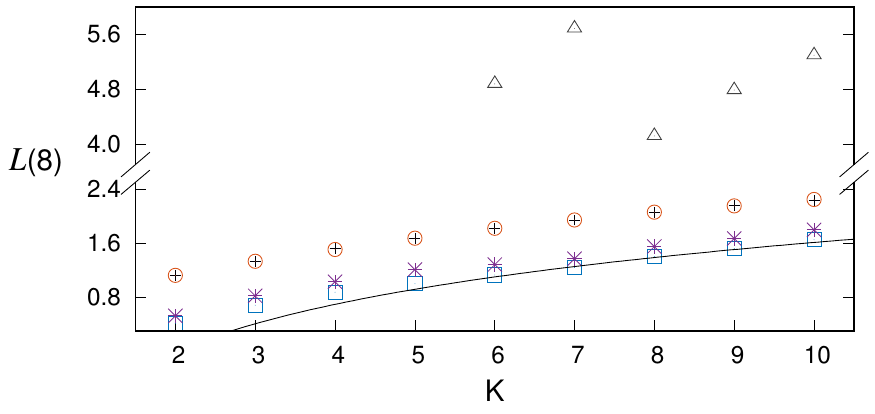}\\
\caption{
Standard map: generalized exponent $L(8)$ versus map parameter $\mathsf{K}$. 
Shown are the results of imperfect cloning (circles),
noisy dynamics plus perfect cloning (crosses),
brute-force method (stars),
and Gaussian approximation (triangles).
Numerical results for the Lyapunov exponent $\lambda$ are represented by squares (numerical) 
and compared to Chirikov's analytical approximation [Eq.~\ref{eq:lambda_app}, full line]. 
All values were obtained for 
time $N=320$, 
sample size $K=10^3$, and 
noise amplitude $\eta=10^{-5}$.
}
\label{fig:std_vs_k}
\end{figure}

\begin{figure}[ht]
\centering
\includegraphics[scale=0.95]{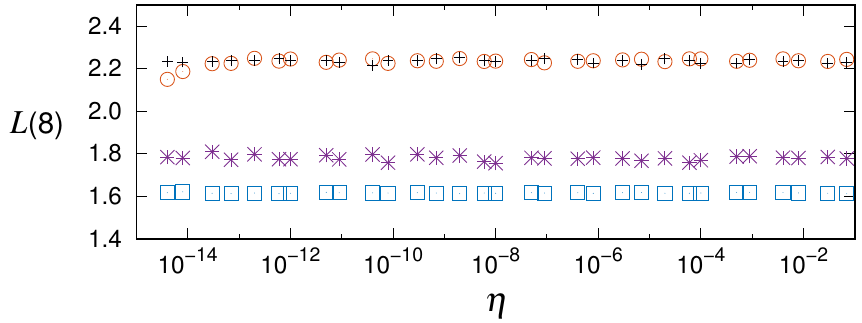}\\
\caption{
Standard map: generalized exponent $L(8)$ versus noise amplitude $\eta$. 
Shown are the results of imperfect cloning (circles), 
noisy dynamics plus perfect cloning (crosses),
brute-force method (stars), and 
Lyapunov exponent $\lambda$ (squares).
Sample size is $K=10^3$, time $N=320$, noise amplitude $\eta=10^{-5}$, 
and map parameter $\mathsf{K}=10$. 
}
\label{fig:std_k10_cloning}
\end{figure}

Thus, we will compare the method that uses imperfect cloning with 
the standard method (noisy dynamics plus perfect cloning), taking as reference the
results of the Gaussian approximation and brute-force sampling.
Figure~\ref{fig:std_vs_k} shows such a comparison as a function of the map parameter $\mathsf{K}$.
Figure~\ref{fig:std_k10_cloning} shows results versus noise amplitude $\eta$.
In both cases, the noisy dynamics was obtained by adding noise only to $p$.

The results for the standard are very similar to those for the tent map.
Both importance sampling methods are insensitive to noise amplitude and
produce almost identical results for $L(8)$, 
while brute-force sampling yields too low values.

The Gaussian approximation deserves a separate comment. 
This approximation relies on the cumulants $\kappa_1$ and $\kappa_2$ of $P(\lambda)$,
i.e., mean and variance (\ref{eq:Lgauss}). 
Let us describe how $P(\lambda)$ evolves as the chaoticity parameter $\mathsf{K}$ increases.
Consider, for instance, $\mathsf{K} \approx 2$, when phase space is covered almost equally by 
regular islands and a chaotic sea \cite{manchein12}.
If we choose initial conditions uniformly distributed over phase space,
then we obtain a bimodal $P(\lambda)$, 
showing one peak near the value of $\langle \lambda \rangle$ corresponding to the chaotic sea, 
and other one at $\lambda=0$ (associated to regular regions).
As time grows both peaks become thinner, but their position do not change significantly.
Because of this, the variance remains finite, and the Gaussian approximation to $L(q)$ diverges
linearly with time.

However, even if we launch the trajectories from the chaotic sea, a bimodal $P(\lambda)$ results 
\cite{sepulveda89,szezech05}.
In this case, the secondary peak at $\lambda=0$ is due to long-time trapping at structures 
surrounding the islands \cite{tomsovic07}.
As the map becomes more chaotic, i.e., for larger $\mathsf{K}$, the peak at $\lambda=0$ vanishes
and $P(\lambda)$ tends to a unimodal distribution 
(see, e.g., the figure for $\mathsf{K}=6$ in \cite{szezech05}).
For $\mathsf{K} \gtrsim 5$ the variance decays like $1/t$:
\begin{equation}
\kappa_2 \, t = C(\mathsf{K})  \, ,
\label{eq:kappa2}
\end{equation}
with $C(\mathsf{K})$ a highly oscillating function \cite{tomsovic07}.
This explains the fluctuations of $L_G(8)$ seen in Fig.~\ref{fig:std_vs_k}.
For each (integer) value of $\mathsf{K}$ we considered only one very long ($N=10^6$) trajectory 
in the chaotic sea divided into 1000 segments of length 1000.
For $\mathsf{K} \le 5$ we did not observe convergence of the Gaussian approximation up to the considered times. 
Even when $L_G$ converges it does to values higher than those calculated with the importance sampling algorithms.
This a clear manifestation of intermittent motion, in the present case, caused by a mixture of chaos and regularity. 

Incidentally, the importance sampling algorithms are insensitive to trapping by regular structures because
they clone those trajectories having Lyapunov exponents larger than the average (for $q>0$) and prune those
having $\lambda \approx 0$ \cite{castiglione08}.

%%%%%%%%%%%%%%%%%%%%%%%%%%%%%%%%%%%%%%%%%%%		
\section{Coupled  maps} 
\label{sec:coupledmaps}		

Crisanti, Paladin, and Vulpiani (CPV) studied a ring of coupled symplectic maps 
defined as follows: \cite{crisanti88}
\begin{eqnarray}
   q^{(i)}_{n+1}    &=&     q^{(i)}_{n}   +  p^{(i)}_{n}    \,, \\ \nonumber
   p^{(i)}_{n+1}    &=&     p^{(i)}_{n}   +  
	                               \epsilon \{ g[q^{(i+1)}_{n+1} - q^{(i  )}_{n+1} ]
	                                         - g[q^{(i  )}_{n+1} - q^{(i-1)}_{n+1} ]\}   
\label{eq:CPV}
\end{eqnarray}
with $1 \le i \le D$ and coupling function $g(x)=\epsilon \sin^\beta(x)$, where $\beta$ is an odd integer.
All variables are taken $\mod \, 2\pi$ (see also \cite{falcioni91,kuptsov11}).
They were interested in the question: does intermittency disappear in the thermodynamic 
limit, i.e., as $D \to \infty$? 
They used $L(q)$ as the quantifier of intermittency, 
and calculated numerically $L(0)=\lambda$, $L(1)$ and $L(2)$
for $D$ up to 80, and several values of $\beta$ and $\epsilon$.

We will compare (some of) their results with ours, obtained 
using importance sampling, either with dynamic noise or imperfect cloning.
(They do not describe their numerical method \cite{crisanti88}.)
We used additive noise in each $p^{(i)}$ for both importance sampling schemes. 	
Table~\ref{tab1} displays the results.
\begin{table}[ht]
\vspace{1pc}
\begin{center}
\begin{tabular}{|c|c|c|c|c|c|c|c|c|}  
\hline
  ~$\beta$~    &
  ~$\epsilon$~ &
	~$D$~        &
  ~$q$~        &  
  ~CPV~        &  
  ~~B~~        & 
  ~IC~         & 
  ~DN~         &
  ~GA~
\\ \hline \hline
  &     &    & 0  & 0.676[5]    & 0.674[1]   & --         & --        & --         \\ \cline{4-9}   % sigma=0.020
1 & 1.0 & 5  & 1  & 0.766[5]    & --         & 0.772[1]   & 0.774[1]  & 0.86[2]    \\ \cline{4-9}   
  &     &    & 2  & 0.856[5]    & --         & 0.867[1]   & 0.863[1]  & 1.020[3]   \\ \hline
	&     &    & 0  & 0.723[5]    & 0.720[1]   & --         & --        & --         \\ \cline{4-9}   % sigma=0.013
1 & 1.0 & 10 & 1  & 0.793[5]    & --         & 0.795[1]   & 0.795[1]  & 0.810[2]   \\ \cline{4-9}
  &     &    & 2  & 0.856[5]    & --         & 0.874[1]   & 0.872[1]  & 0.883[3]   \\ \hline
	&     &    & 0  & 0.397[2]    & 0.362[1]   &  --        & --        & --         \\ \cline{4-9}   % sigma=0.017
3 & 0.4 & 5  & 1  & 0.437[5]    & --         & 0.417[1]   & 0.420[1]  & 0.496[4]   \\ \cline{4-9}
  &     &    & 2  & 0.473[5]    & --         & 0.476[1]   & 0.476[1]  & 0.652[3]   \\ \hline
	&     &    & 0  & 0.372[2]    & 0.392[1]   & --         & --        & --         \\ \cline{4-9}   % sigma=0.012
3 & 0.4 & 10 & 1  & 0.434[5]    & --         & 0.438[1]   & 0.436[1]  & 0.453[4]   \\ \cline{4-9}
  &     &    & 2  & 0.478[7]    & --         & 0.486[1]   & 0.484[1]  & 0.526[3]   \\ \hline
   &      &   & 0  & 0.027[1]  & 0.0169[1]  &  --       & --       & --         \\ \cline{4-9}  % sigma=0.012 
5* & 0.02 & 5 & 1  & 0.051[1]  & --         & 0.069[2]  & 0.068[1] &  $\times$  \\ \cline{4-9}
	 &      &   & 2  & 0.080[1]  & --         & 0.093[1]  & 0.096[2] &  $\times$  \\ \hline
	 &      &    & 0  & 0.032[1] & 0.0250[1]  & --        & --       & --         \\ \cline{4-9}  % sigma=0.012 
5* & 0.02 & 10 & 1  & 0.061[1] & --         & 0.068[1]  & 0.068[1] &  $\times$  \\ \cline{4-9}
	 &      &    & 2  & 0.091[1] & --         & 0.093[1]  & 0.096[1] &  $\times$  \\ \hline
	&     &    & 0  & --   & 0.777[1]   &  --        & --        & --             \\ \cline{4-9}   % sigma=0.016
5 & 1.0 & 5  & 1  & --   & --         & 0.876[1]   & 0.873[2]  &  0.899[5]      \\ \cline{4-9}
	&     &    & 2  & --   & --         & 0.961[2]   & 0.965[2]  &  1.012[5]      \\ \hline
	&     &    & 0  & --   & 0.820[1]   & --         & --        & --             \\ \cline{4-9}   % sigma=0.012
5 & 1.0 & 10 & 1  & --   & --         & 0.894[1]   & 0.892[1]  &  0.890[2]      \\ \cline{4-9}
  &     &    & 2  & --   & --         & 0.973[1]   & 0.969[1]  &  0.956[4]      \\ \hline
\end{tabular}
\end{center}
\caption{%
CPV are numerical results by Crisanti, Paladin, and Vulpiani \cite{crisanti88}.
B=Benettin's method; 
IC=imperfect cloning; 
DN=dynamical noise + perfect cloning; 
GA=Gaussian approximation. 
In all cases 
$\eta=10^{-5}$ (noise amplitude), $N=K=10^3$; except in (*): $N=K=10^4$. 
The square brackets contain the estimated error in the least significant figure 
(e.g., the notation $0.82[2]$ stands for $0.82\pm 0.02$).
Crosses indicate that the Gaussian approximation failed to converge.}
\label{tab1}
\end{table}
First of all we must highlight the coincidence of both importance-sampling results for all cases, i.e.,``IC=DN". 
Second: these results are consistent with CPV for the strong-coupling cases $\epsilon=1.0$ and $\epsilon=0.4$.

The cases having $\epsilon=0.02$ correspond to weakly coupled maps and exhibit several anomalies.
To start with, the CPV Lyapunov exponents are very different from ours 
(we conjecture that this may be associated to different sampling schemes).
Also their generalized exponents $L(1)$ and $L(2)$ do not coincide with ours, 
though in this case the difference is only 10/20\%.
Finally, the Gaussian approximation diverges with time.

Falcioni et al \cite{falcioni91} also studied the present system (\ref{eq:CPV}) for $\beta=1$.
They observed that the numerical $P(\lambda)$, 
obtained following many trajectories starting from different initial conditions, 
has a finite variance $\kappa_2$.
They stress that even when the chaotic regions have very small probability
most trajectories have a positive $\lambda$, even if the
values of the $\lambda$ depend on the initial conditions.
However, for small values of the coupling constant the tendency to a unique chaotic phase is very slow.
This explains the finiteness of $\kappa_2$ and, consequently, 
the failure of the Gaussian approximation for the case $(\beta=5,\epsilon=0.02)$.

In order to verify that this anomalous behavior is due to weak coupling, 
we analized the case $(\beta=5,\epsilon=1.0)$. 
Here we verified that $\kappa_2 \, t$ tends to a definite value, like in the other cases of
strong coupling depicted in Table~\ref{tab1}. 
Indeed, for $D=10$ the Gaussian approximation works very well.
This is consistent with the approximate linearity of $L(q)$ 
as inferred from the numerical data from columns ``B" and ``IC", that is,
$[L(2)-L(1)]/[L(1)-L(0)] \approx 1$ (=1.05). 
For the sake of completeness we list all the eight ``linearity quotients", 
corresponding to eight cases appearing in Table~\ref{tab1} (from top to bottom):
\{0.97,     1.05,    1.07,    1.04,    0.46,     0.58,     0.86,     1.07\}.
%{0.969388, 1.05333, 1.07273, 1.04348, 0.460653, 0.581395, 0.858586, 1.06757}
%
Not surprisingly the lowest quotients --fifth and sixth-- correspond to smallest coupling, $\epsilon=0.02$,
where $L_G$ is not even defined.

%%%%%%%%%%%%%%%%%%%%%%%%%%%%%%%%%%%%%%%%%%%		
\section{Conclusions} 
\label{sec:conclusions}		

We put forward and tested a novel importance-sampling algorithm
for calculating Lyapunov generalized exponents of deterministic systems.
The algorithm modifies 
Tailleur-Kurchan's and Vanneste's cloning/pruning methods
by introducing imperfect cloning.
This avoids the standard procedure of adding noise to dynamics, 
thus preserving the simplicity of the equations of motion. 
Moreover, in Hamiltonian systems, energy conservation is easily imposed in our algorithm: 
we just renormalize momenta after cloning. 
This contrasts with the use of relatively sophisticated algorithms 
for implementing energy-conserving noisy dynamics \cite{tailleur07}.

We showed that our algorithm performs as well as the standard method \cite{tailleur07,vanneste10},
provided that the parameters (number of trajectories, propagation time, resampling frequency, noise amplitude) 
are properly chosen. 
Curiously enough, both importance-sampling methods are insensitive to noise amplitude
(at least for the considered systems, and noise level within certain bounds), 
thus the limit $\eta \to 0$ is unnecessary --it suffices to fix $\eta$ to a convenient value.

We believe that this method is an important contribution to the suite of tools for computing 
$L(q)$ 
\cite{crisanti93,vanneste10,tailleur07,pikovsky16}, 
especially for high dimensional Hamiltonian systems.
Imperfect cloning could also be used in Lyapunov weighted dynamics, 
designed to locate special structures in Hamiltonian systems, e.g., 
small islands of regularity, Arnold webs, separatrices, etc., which are characterized by 
a Lyapunov exponent off the average value \cite{tailleur07}. 
In this way, these importance-sampling methods are complementary to those developed by 
Manchein et al.~\cite{manchein12,manchein14} and da Silva et al.~\cite{dasilva15}
for the characterization of weak chaos in high-dimensional Hamiltonian systems.

{\bf Acknowledgments:} 
We acknowledge Brazilian agencies CNPq and FAPERJ for partial financial support.

\end{document}